# Influence of annealing temperature on the structural, topographical and optical properties of sol–gel derived ZnO thin films

Joydip Sengupta [a,*], R.K. Sahoo [b], K.K. Bardhan [a], C.D. Mukherjee [a]

[a] Experimental Condensed Matter Physics Division, Saha Institute of Nuclear Physics, 1/AF Bidhannagar, Calcutta 700064, India
[b] Materials Science Centre, Indian Institute of Technology, Kharagpur 721302, India

## ABSTRACT

This investigation deals with the effect of annealing temperature on the structural, topographical and optical properties of Zinc Oxide thin films prepared by sol–gel method. The structural properties were studied using X-ray diffraction and the recorded patterns indicated that all the films had a preferred orientation along (002) plane and the crystallinity along with the grain size were augmented with annealing temperature. The topographical modification of the films due to heat treatment was probed by atomic force microscopy which revealed that annealing roughened the surface of the film. The optical properties were examined by a UV–visible spectrophotometer which exhibited that maximum transmittance reached nearly 90% and it diminished with increasing annealing temperature.

## 1. Introduction

Zinc Oxide (ZnO) is a II–VI semiconductor with hexagonal wurtzite crystal structure. Owing to the direct band gap (3.3 eV), large exciton binding energy (60 meV), piezoelectric properties and high transmittance in the visible region; ZnO has become a potential candidate for variety of applications e.g. short-wavelength light emitting devices [1], solar cell [2], surface acoustic wave devices [3] etc. Beside the numerous potential applicability of ZnO the additional advantages are non-toxicity, chemical and thermal stability, ready availability and low cost. Thus, in recent years ZnO has received considerable attention towards the fabrication of good quality thin films, employing pulsed laser deposition, RF magnetron sputtering, electrodeposition, chemical vapor deposition, spray pyrolysis, sol–gel process etc. Among different synthesis methods, the sol–gel process of film preparation is attractive as it provides simple, inexpensive preparation of a large-area homogeneous thin film along with excellent compositional control, lower crystallization temperature and uniform film thickness.

The principal factors affecting the microstructure and properties of the sol–gel derived film are aging time of sol [4], thickness of the film [5], annealing treatment [6] etc. In particular, the annealing treatment is a primary factor which significantly affects the physical properties of sol–gel derived films. Therefore, it is necessary to systematically investigate the effect of annealing on structural, topographical and optical properties of the sol–gel derived ZnO films, which are important parameters to be taken under consideration for optoelectronic applications of these films.

## 2. Experimental procedure

Zinc acetate dihydrate ($Zn(CH_3COO)_2 \cdot 2H_2O$), isopropyl alcohol (IPA) and diethanolamine ($C_4H_{11}NO_2$) were used as starting material, solvent and stabilizer, respectively. First, zinc acetate dihydrate was dissolved in IPA and then diethanolamine was slowly added into the solution under magnetic stirring. The molar ratio of diethanolamine to zinc acetate was kept at 1. The resulting mixture was stirred for 1 h at 65 °C, and then 3 h at room temperature to yield a clear and homogeneous 0.5 M solution. The solution was aged for 48 h at room temperature. Thin films of ZnO were prepared by spin coating the aged solution onto pre-cleaned quartz substrates at rotation speed of 3000 rpm for 30 s in ambient condition. Afterwards the films were dried at 300 °C for 10 min in air to evaporate the solvent and organic residues. The ZnO films were then inserted into a furnace and annealed in air at 400, 550 and 700 °C for 1 h.

An atomic force microscope (AFM) (Nanonics Multiview 4000™) in intermittent contact mode, Philips X-ray diffractometer (XRD) (PW1729) with Co source and UV–visible (UV–vis) spectrophotometer (PerkinElmer, Lambda 35) were used to characterize the annealed ZnO films.

* Corresponding author. Fax: +91 33 2337 4637.
E-mail address: joydipdhruba@gmail.com (J. Sengupta).

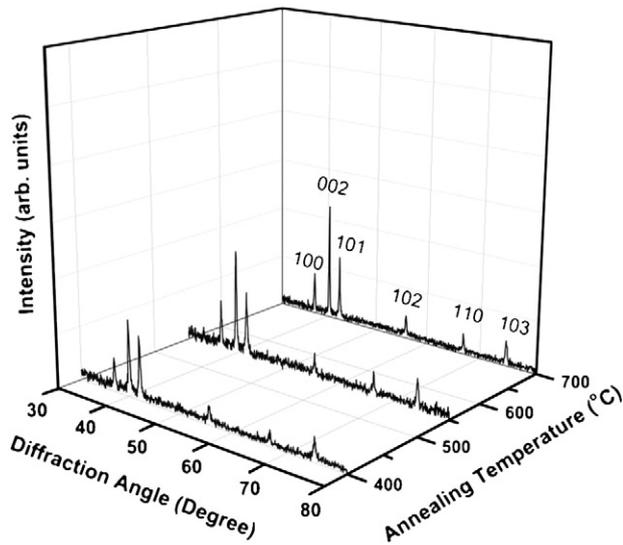

**Fig. 1.** X-ray diffraction spectra of annealed ZnO thin films deposited on quartz substrate using spin coating.

## 3. Results and discussion

The XRD patterns of ZnO thin films annealed at three different temperatures (Fig. 1) showed that all annealed films were polycrystalline in nature with a hexagonal wurzite structure. The diffractograms also revealed that preferred orientation along (002) was common in all the films and the intensity of (002) peak gradually increased with the increasing annealing temperature. It was reported that the preferential crystal orientation of ZnO films has a profound impact on ZnO-based device properties [7]. Therefore, in order to precisely investigate the effect of annealing temperature on the degree of orientation of the (002) plane a formula proposed by Lotgering was used [8].

$$F(hkl) = \frac{P(hkl) - P_0(hkl)}{1 - P_0(hkl)} \quad (1)$$

where $F(hkl)$ is the degree of $(hkl)$ orientation, $P(hkl) = I(hkl)/\sum I(hkl)$ and $P_0(hkl) = I_0(hkl)/\sum I_0(hkl)$. Here $I(hkl)$ is the $(hkl)$ peak intensity and $\sum I(hkl)$ is the sum of the intensities of all peaks in the ZnO films' diffraction data. $I_0(hkl)$ is the $(hkl)$ peak intensity and $\sum I_0(hkl)$ is the sum of the intensities of diffraction peaks in the reference data (JCPDS 36–1451). The values of degree of orientation of the (002) plane of the annealed ZnO films (Table 1) revealed that the preferential c-axis orientation perpendicular to the substrate surface augmented with the increase in annealing temperature. The increment of degree of orientation could be explained as, with the increase of annealing temperature ZnO crystallites gain enough energy and orient themselves along (002) plane as it possesses highest atomic packing density and minimum surface energy [9].

The average grain size ($D$) of the ZnO films was also calculated using the full width at half maximum (FWHM) of (002) peak from the Scherrer's equation

$$D = \frac{K\lambda}{\beta \cos\theta} \quad (2)$$

where $K = 0.9$ is the shape factor, $\lambda$ is the wavelength of incident X-ray, $\beta$ is the FWHM measured in radians and $\theta$ is the Bragg angle of diffraction peak. It was observed (Table 1) that, as the annealing temperature increased from 400 to 700 °C, the FWHM value exhibited a tendency to decrease. The trend of FWHM values implied that the crystallinity of the ZnO thin films was improved as the annealing temperature was increased [10]. It was also observed (Table 1) that average grain size was increased with increasing annealing temperature. This could be explained by considering the thermal annealing induced coalescence of small grains by grain boundary diffusion which caused major grain growth [10].

Three-dimensional AFM images as given in Fig. 2 showed the influence of post-growth annealing on the surface morphology of ZnO thin films. All the samples showed good homogeneity and no cracks were noted. Upon close inspection of the AFM images, it was observed that the grain sizes become larger with the increase of annealing temperature. At high temperatures, atoms had enough diffusion activation energy to occupy the energetically favorable site in the crystal lattice and eventually grains with the lower surface energy became larger, which was consistent with the results of XRD. The root mean square (RMS) roughness of the ZnO thin films annealed at 400, 550 and 700 °C was listed in Table 1, which revealed that the RMS roughness value of the annealed ZnO films was also increased with the increasing annealing temperature. This could be explained in terms of major grain growth which yields an increase in the surface roughness [6].

Optical properties of ZnO films deposited on quartz substrate and annealed at 400, 550 and 700 °C were analyzed by spectrophotometric measurements. The optical transmittance spectra of annealed ZnO films (Fig. 3) exhibited sharp absorption edges in the wavelength region around 380 nm. The average transmittance of the annealed ZnO thin films (Table 1) indicated that the transmittance had decreased with the increase in annealing temperature. It was previously reported that surface roughness strongly affects the transparency of ZnO-based thin films [11]. The AFM measurements already revealed that the roughness values of the ZnO films had increased with annealing temperature. So it could be stated that the major reason for the decrease in transmittance with higher annealing temperature might be due to the rough surface scattered and reflected light.

From transmittance measurements the optical band gap could also be estimated by employing the Tauc model:

$$(\alpha h\nu) = A(h\nu - E_g)^{1/2} \quad (3)$$

where $\alpha$ is the absorption coefficient, $h\nu$ is the photon energy, $A$ is a constant and $E_g$ is the optical bandgap. The optical bandgap of ZnO thin films annealed at different temperatures was determined by extrapolation of the straight section to the energy axis of the plot of $(\alpha h\nu)^2$ versus photon energy (inset of Fig. 3). Table 1 lists the extrapolated bandgap values of ZnO thin films which showed decrease in bandgap with increasing annealing temperature. The shifts of the optical band gap might be attributed to the decreased defect of the thin films with the increase in annealing temperature.

**Table 1**
The data evaluated form the XRD, AFM and UV–vis measurements of sol–gel derived ZnO thin films after annealing at different temperatures.

| Annealing temperature (°C) | Degree of orientation of (002) plane | FWHM of (002) plane (degree) | Average grain size (nm) | RMS roughness (nm) | Average transmittance[a](%) | Optical bandgap (eV) |
|---|---|---|---|---|---|---|
| 400 | 0.22 | 0.39303 | 25 | 3.38 | 89 | 3.26 |
| 550 | 0.25 | 0.31374 | 31 | 6.33 | 85 | 3.25 |
| 700 | 0.32 | 0.25019 | 39 | 11.16 | 80 | 3.24 |

[a] The average transmittance values were calculated using the transmittance data from 420 to 800 nm.

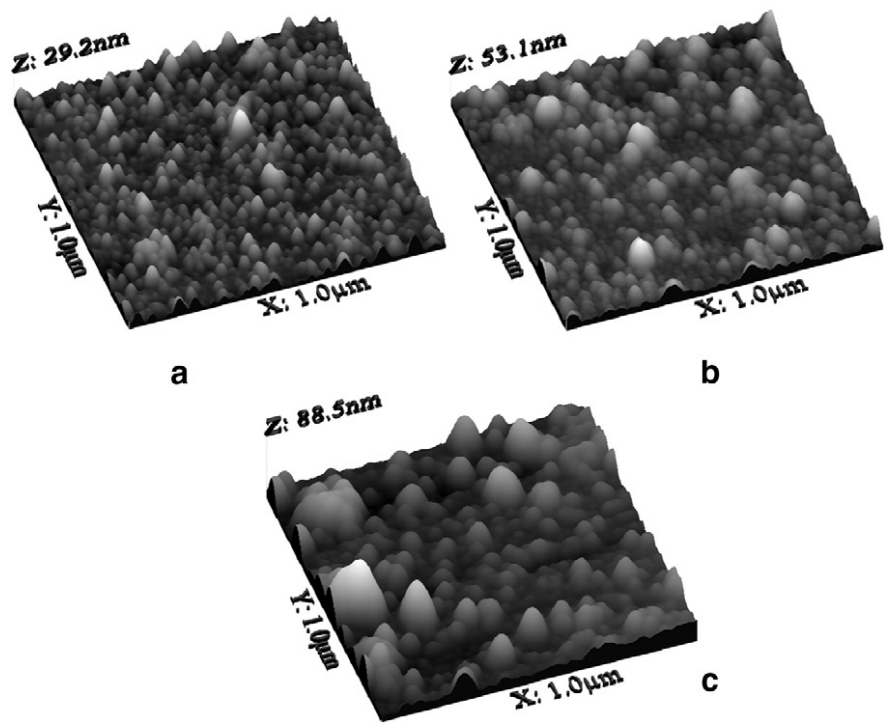

Fig. 2. Three dimensional AFM images of ZnO thin films deposited on quartz substrate after annealing at different temperatures in air for 1 h (a) 400 °C, (b) 550 °C, (c) 700 °C.

## 4. Conclusions

We have investigated the structural, topographical and optical properties of sol–gel derived ZnO thin films deposited on quartz substrate with respect to annealing temperature. The study revealed that all the ZnO thin films had preferred c-axis orientation with hexagonal wurtzite structure. Moreover, the crystallinity, degree of preferred orientation and average grain size of the films increased with annealing temperature. The annealed ZnO films also showed good transparency in the wavelength region of 400 to 800 nm. However, it was observed that the optical transmittance had decreased with increase in annealing temperature which could be explained by the increased surface roughness of the deposited films as confirmed by AFM.

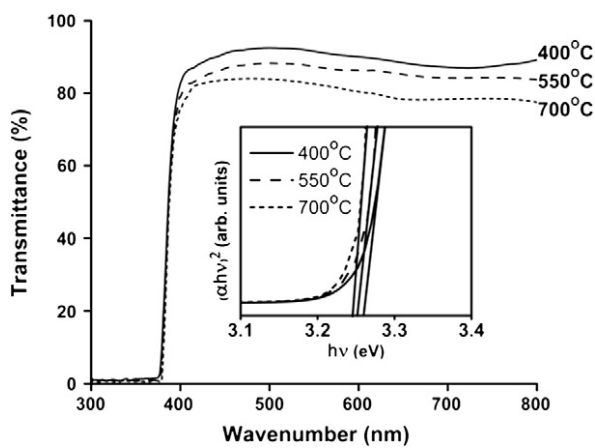

Fig. 3. Optical transmittance spectra of sol–gel derived ZnO thin films after annealing at different temperatures. (Inset) Tauc's plot of annealed ZnO films on quartz substrate.